%% file: main.tex
\documentclass[preliminary,copyright,creativecommons]{eptcs}
\providecommand{\event}{ICLP 2025 Technical Communication} % Name of the event you are submitting to

\usepackage{iftex}

\ifpdf
  \usepackage{underscore}         % Only needed if you use pdflatex.
  \usepackage[T1]{fontenc}        % Recommended with pdflatex
\else
  \usepackage{breakurl}           % Not needed if you use pdflatex only.
\fi

\usepackage{amsmath}
\usepackage{graphicx}
\usepackage{multirow}
\usepackage{alltt}
\usepackage{amssymb}
\usepackage{tcolorbox}
\usepackage{esvect}
\usepackage{url}
\usepackage{plain}
\usepackage{float}

\title{Automated Theorem Proving for Prolog Verification\footnote{This paper is an updated version of \cite{MesnardMP24} that appeared  in the 2024 LPAR Complementary Volume.}
}
\author{Fred Mesnard \and Thierry Marianne \and \'Etienne Payet
\institute{LIM, universit\'e de La R\'eunion, France}
\email{\{frederic.mesnard,thierry.marianne,etienne.payet\}@univ-reunion.fr}}

\newcommand{\titlerunning}{Automated Theorem Proving for Prolog Verification}
\newcommand{\authorrunning}{F. Mesnard, T. Marianne \& \'E. Payet}

\hypersetup{
  bookmarksnumbered,
  pdftitle    = {\titlerunning},
  pdfauthor   = {\authorrunning},
  pdfsubject  = {EPTCS},               
  pdfkeywords = {Prolog, Verification, ATP} % Uncomment and enter keywords specific to your paper
}
\begin{document}

\maketitle

\begin{abstract}
%Summary of the paper in about 250 words.
LPTP (Logic Program Theorem Prover) is an interactive %first-order 
natural-deduction-based theorem 
prover for pure Prolog programs with negation as failure, unification with the occurs check,
and a restricted but extensible set of built-in predicates. 
With LPTP, one can formally prove termination and partial correctness of such Prolog programs.
LPTP was designed in the mid-1990's by Robert F. St{\"a}rk. 
It is written in ISO-Prolog and comes with an Emacs user-interface. 

From a theoretical point of view, in his publications about LPTP,
St{\"a}rk associates a set of first-order axioms IND($P$) to the considered Prolog program $P$. 
IND($P$) contains the Clark's equality theory for $P$,  definitions
of success, failure and termination for each user-defined logic procedure in $P$, 
axioms relating these three points of view, and an axiom schema for 
proving inductive properties. LPTP is thus a dedicated proof editor
where these axioms are hard-wired. 

We propose to translate these axioms as first-order formulas (FOFs), 
and apply automated theorem provers to check the property of interest.
Using  FOF  as an intermediary language, we experiment 
the use of automated theorem provers for Prolog program verification.
We evaluate the approach over  a benchmark of about 400 properties of Prolog programs
from the library available with LPTP.
Both the  compiler which generates a set of FOF files from a given input Prolog program
together with its properties
and the benchmark are publicly available.
\end{abstract}

%\begin{keywords} Prolog  verification, automated theorem proving \end{keywords}

%\newpage
\section{Introduction}
In the mid-1990's, Robert F. St{\"a}rk defined  a framework for Prolog verification  
\cite{Staerk95a,Staerk98a}.
He considered a subset of ISO-Prolog \cite{ISO-Prolog95}: pure Prolog programs with negation as failure, unification with the occurs check,
and allowed a restricted but extensible set of built-in predicates.
He presented a first-order formalisation with axiom schemas of the usual operational semantics 
of Prolog. A safeness condition included in termination condition imposes
groundness before evaluation of negated goals.
He showed soundness and completeness for termination, success, and failure.
The framework also allows partial correctness properties to be proved by induction
w.r.t. the clauses defining predicates, considered as inductive definitions.
Some examples will be discussed in Section \ref{lptp:fof} and Section \ref{experimental:results}.
The  logical theory was hard-wired 
in an interactive  dedicated first-order natural-deduction-based theorem prover called LPTP
(Logic Program Theorem Prover). 
St{\"a}rk  implemented
LPTP in ISO-Prolog, together with an Emacs user-interface,
an HTML and \TeX \ manager, a  detailed user-manual,
and a library of  predicates for Peano numbers, integers, lists,
sorting algorithms, etc. with numerous proven properties.
A copy of LPTP is vailable at \url{https://github.com/FredMesnard/lptp}.

Thirty years later, LPTP is still running on any ISO-Prolog processor, 
with its initial interface. 
Today, formal verification of computer programs is an established 
discipline within computer science. Nonetheless,
program verification by interactive theorem proving is still a slow process
and requires non-trivial skills. 
On the other hand, during the last three decades, the increase in computing power
and the advances in automated theorem proving %theory and technology 
have been notable.
For instance,   TPTP (Thousands of Problems for Theorem Provers, \cite{Sut17})
is a library of test problems for automated theorem proving. It provides online
tools to check the syntax of input problems and apply a bunch of user selected automated theorem provers.
Among them, E \cite{Schulz19a} and Vampire \cite{KovacsV13} are two powerful freely available
automated theorem provers, performing very well in many international competitions over the years.
Interactive theorem prover implementers were able 
to take advantage of these progress by implementing so-called \emph{hammers} 
for their tools, see e.g.  \cite{PaulsonB10,BlanchetteKPU16}.

This evolution raises the following questions: 
can we also use the TPTP FOF \emph{Esperanto} to formulate
the logic theory St{\"a}rk associates to a logic program? 
Can we use \emph{off-the-shelf} TPTP provers and obtain automatic proofs in 
reasonable time? Can we get  an acceptable success rate with such an approach?

The main contribution of this paper is the following.
Using  FOF (\emph{First-Order Form}, one of the logic languages proposed by TPTP, see \cite{Sut23})
as an intermediary language, we describe the first -- to the best of our knowledge --
experiment of the use of automated theorem provers, namely E and Vampire,
for Prolog program verification, including termination and partial correctness. 
We evaluate the approach over about 400 properties of Prolog programs.
Both the compiler applying St{\"a}rk's theory to a given input Prolog program and its properties
to a set of FOF files 
and the benchmark are publicly available at {\url{https://github.com/atp-lptp/automated-theorem-proving-for-prolog-verification}}.

We organize the paper as follows.
The next section presents a brief summary of the LPTP system.
The third section describes step by step how to compile
a Prolog program, its associated LPTP axioms and a property of interest 
into a FOF file.
%Section 4 studies the logical relationship between the LPTP axioms 
%and  the FOF compiled form.
Then we present an experimental evaluation, related work and we conclude.

\section{Notation}
\label{notation}

FOF (\emph{First Order Form}) is a well-known logic language from TPTP
for expressing first-order logic (FOL) axioms and conjectures.
A formula is written 
\texttt{fof(\textit{name},\textit{role},\textit{formula})},
where \textit{name} is the name of the formula, \textit{role} is either \texttt{axiom} or \texttt{conjecture}
and \textit{formula} is informally  defined as follows:
\begin{center}
\begin{tabular}{| l | l | | l | l |}
\hline
FOL & FOF & FOL & FOF  \\
\hline
$A \land B$ &  \texttt{A  \& B} & $\neg p(x)$ & \texttt{\~ \ p(X)} \\
$A \lor B$ & \texttt{A | B} & $\exists x . A$ & \texttt{?[X] : A}\\
$A \rightarrow B$ & \texttt{A => B} & $\forall x . A$ & \texttt{![X] : A}\\
\hline
\end{tabular}
\end{center}

%%\begin{table}%[h!]
%%\centering
%%\caption{An Informal Summary of the FOF Syntax}
%%{\tablefont\begin{tabular}{@{\extracolsep{\fill}}ccccc}
%\begin{tabular}{| l | l | | l | l |}
%%\topline
%FOL & FOF &  & FOL & FOF  
%%\midline
%$A \land B$ &  \texttt{A  \& B} &  & $\neg p(x)$ & \texttt{\~ \ p(X)} \\
%$A \lor B$ & \texttt{A | B} &  & $\exists x . A$ & \texttt{?[X] : A}\\
%$A \rightarrow B$ & \texttt{A => B} &  & $\forall x . A$ & \texttt{![X] : A}
%%\botline
%\end{tabular}
%%\end{tabular}}
%%\end{table}

Numerous examples will appear in the next sections.%, sometimes in an extended intuitive syntax.

Let $P$ be a pure logic program where negative literals
may appear in the body of clauses (also called \emph{normal  program} in  \cite{Lloyd87a}).
For sake of conciseness,
we do not consider built-in predicates (see \cite{Staerk98a}
for a full treatment) other than the equality \texttt{=/2}.
We start with $\cal{L}$, the first-order language associated with $P$. 
The \emph{goals} of $\cal{L}$ are:
$$G,H ::= \texttt{true} \ | \texttt{fail}\ | \ s = t \ | \ A \ | \ \texttt{\textbackslash +} \ G \ | 
     \ (G,H) \  | \ (G;H) \ | \ \texttt{some} \ x \ G$$
where $s$ and $t$ are two terms, $x$ is a variable and $A$ is an atomic goal. The goals of $\cal{L}$ have the operational semantics specified 
by ISO-Prolog \cite{ISO-Prolog95} assuming the occurs check.

$\hat{\cal{L}}$ is the specification language of LPTP.
For each user-defined predicate symbol $R$,  $\hat{\cal{L}}$ does not include $R$, but instead it contains
three predicate symbols $R^s$, $R^f$, $R^t$ of the same arity as $R$,
which respectively express success, failure and termination of $R$. 
$\hat{\cal{L}}$ also contains a unary  constraint for groundness $gr$, expressing that its argument is ground.
The \emph{formulas} of $\hat{\cal{L}}$ are:
$$\phi,  \psi ::= \top \ | \ \bot \ | \ s = t  \ | \ R(\vv{t}) \ | 
  \  \neg \phi \ |  \ \phi \land \psi \ |  \ \phi \lor \psi \ |  \ \phi \rightarrow \psi \ | \ \forall x \phi \ | \ \exists x \phi $$
where $\vv{t}$ is a sequence of  $n$ terms and $R$ denotes a $n$-ary predicate symbol of $\hat{\cal{L}}$. The semantics of $\hat{\cal{L}}$ 
is the first-order predicate calculus of classical logic.

For any of the user-defined  logic procedure $R$ in a logic program $P$,
$D^P_R(\vv{x})$ denotes its Clark's \emph{if-and-only-if} completed definition, cf. \cite{Cla78, Lloyd87a}.

For defining the declarative semantics of logic programs, 
St{\"a}rk uses three syntactic operators \textbf{S}, \textbf{F} and \textbf{T} which map
goals of $\cal{L}$ into $\hat{\cal{L}}$-formulas. Intuitively, $\textbf{S}G$ means $G$ succeeds
(any breadth-first evaluation of $G$ succeeds), 
$\textbf{F}G$ means $G$  fails (the ISO-Prolog evaluation stops without any answer), 
and $ \textbf{T}G$ means $G$ terminates
(the ISO-Prolog evaluation produces a finite number of answers then stops).
The definition of the operators follows:
\begin{flushleft}
\small
\begin{tabular}{l l l l}
%\hline
%%
 $\textbf{S}R(\vv{t}) := R^s(\vv{t})$ 	&   
 	$\textbf{S}  \ \texttt{true} := \top$ & $\textbf{S} \ \texttt{fail} := \bot$ & $\textbf{S} (s = t) := (s = t)$ \\
 $\textbf{S}  \texttt{\textbackslash +} G := \textbf{F} G$ & $\textbf{S}( G, H) := \textbf{S}G \land \textbf{S}H$  &
 	$\textbf{S}( G; H) := \textbf{S}G \lor \textbf{S}H$  & $\textbf{S}(\texttt{some} \ x \ G) := \exists x \textbf{S}G$  \\
	& & & \\
 $\textbf{F}R(\vv{t}) := R^f(\vv{t})$ 	&   
 	$\textbf{F}  \ \texttt{true} := \bot$ & $\textbf{F} \ \texttt{fail} := \top$ & $\textbf{F} (s = t) := \neg (s = t)$ \\
 $\textbf{F}  \texttt{\textbackslash +} G := \textbf{S} G$ & $\textbf{F}( G, H) := \textbf{F}G \lor \textbf{F}H$  &
 	$\textbf{F}( G; H) := \textbf{F}G \land \textbf{F}H$  & $\textbf{F}(\texttt{some} \ x \ G) := \forall x \textbf{F}G$  \\
%%
%	& & & \\
% $\textbf{T}R(\vv{t}) := R^t(\vv{t})$ 	&   
 %	$\textbf{T}  \ \texttt{true} := \top$ & $\textbf{T} \ \texttt{fail} := \top$ & $\textbf{T} (s = t) := \top$ \\
%%
 %$\textbf{T}  \texttt{\textbackslash +} G := \textbf{T} G\land  gr(G)$ & &
 %	$\textbf{T}( G, H) := \textbf{T}G \land( \textbf{F}G \lor \textbf{T}H)$  & \\
% 	$\textbf{F}( G; H) := \textbf{T}G \land \textbf{T}H$  & $\textbf{T}(\texttt{some} X \ G) := \forall x \textbf{T}G$  \\
%%	
%\hline
\end{tabular}
\end{flushleft}

\vspace{0.1cm}
\begin{flushleft}
\small
\begin{tabular}{l l }
%\hline
 $\textbf{T}R(\vv{t}) := R^t(\vv{t})$ 	&   
 	$\textbf{T}  \ \texttt{true} := \top$  \\
$\textbf{T} \ \texttt{fail} := \top$ &  $\textbf{T} (s = t) := \top$  \\
 $\textbf{T}  \texttt{\textbackslash +} G := \textbf{T} G\land  gr(G)$ & 
 	$\textbf{T}( G, H) := \textbf{T}G \land( \textbf{F}G \lor \textbf{T}H)$   \\
$\textbf{T}( G; H) := \textbf{T}G \land \textbf{T}H$  &  $\textbf{T}(\texttt{some} \ x \ G) := \forall x \textbf{T}G$   \\
%%	
%\hline
\end{tabular}
\end{flushleft}

Note that termination requires a safe use of negation, see the definition of 
$\textbf{T}  \texttt{\textbackslash +} G$ where the goal $G$ has to be proved 
terminating and ground at proof time.
Finally, we add the definition of $gr$, which belongs to the specification language and is  needed 
for defining $\textbf{T}  \texttt{\textbackslash +} G$:
\begin{flushleft}
\small
\begin{tabular}{l l }
%\hline
  $gr(\texttt{true}) := \top$ & $gr((G,H)):= gr(G) \land gr(H)$ \\ 	
  $gr(\texttt{fail}) := \top$ & $gr((G;H)):= gr(G) \land gr(H)$ \\ 	
  $gr( s = t) := gr(s) \land gr(t)$ & $gr(\texttt{some} \ x \ G ) := \exists x \ gr(G)$ \\
  $gr(R(t_1, \ldots,t_n)) := gr(t_1) \land \ldots \land gr(t_n)$ & $gr (\texttt{\textbackslash +} G) := gr(G)$\\
%\hline
\end{tabular}
\end{flushleft}

We refer the reader to the papers of St{\"a}rk \cite{Staerk95a,Staerk96a,Staerk97,Staerk98a}
for a complete presentation of LPTP. 

\section{Compiling LPTP axioms to FOF}
\label{lptp:fof}

With LPTP, we prove properties of a logic program $P$ w.r.t. its \emph{inductive extension} IND($P$)
which includes Clark's completion \cite{Cla78} and induction along the definition
of the predicates. St{\"a}rk shows that the inductive extension is always consistent
and proves various correctness and completeness results w.r.t. the operational semantics of Prolog \cite{Staerk98a}.
The first-order theory IND($P$) (cf. \cite{Staerk98a}, pp. 253--254) is defined by 
nine axiom schemas
which we describe now, along with their translation in FOF. 
We omit the fixed point axioms for builtins.  
Let us also point out that the specification language $\hat{\cal{L}}$ of LPTP can
be extended by new function and predicate symbols at a logical level
(there is no associated Prolog code).
As shown in \cite{MesnardMP24},
such function and predicate definitions can also be compiled
into FOF.

\subsection{First steps}
Let us consider the following logic program ADD as our running example.
%\begin{tcolorbox}
\begin{verbatim}
nat(0).                           add(0,Y,Y).
nat(s(X)) :- nat(X).              add(s(X),Y,s(Z)) :- add(X,Y,Z).
\end{verbatim}
%\end{tcolorbox}

We discuss the axioms proposed by St{\"a}rk  and apply them to the ADD program.
\begin{tcolorbox}[coltitle=black!75!black, colbacktitle=black!10!white,
                title=The axioms of Clark's equality theory]
\begin{itemize}
    \item[1.] $f(x_1,\ldots, x_n) = f(y_1,\ldots, y_n) \rightarrow x_i = y_i$ [if $f$ is $n$-ary and $1 \leq i \leq n$]
    \item[2.] $f(x_1,\ldots, x_n) \neq g(y_1,\ldots, y_m)$ [if $n \neq m$ or $f  \not\equiv g$]
    \item[3.] $t \neq x$ [if $x$ occurs in $t$ and $t  \not\equiv x$]
\end{itemize}
\end{tcolorbox}
The first two axioms specify some properties of the
trees built from the  function symbols extracted from the program under consideration. 
The third axiom forbids infinite %rational 
trees.
Note that it is actually an axiom schema, i.e., an infinite set of first order
axioms. We will omit it but we stay sound.
Here is the FOF version (see Section \ref{notation} for the FOF syntax):
\begin{verbatim}
fof(id1,axiom,! [Xx4] : ! [Xx5] : (s(Xx4) = s(Xx5) => Xx4 = Xx5)).
fof(id2,axiom,! [Xx3] : ~ ('0' = s(Xx3))).
\end{verbatim}
\begin{tcolorbox}[coltitle=black!75!black, colbacktitle=black!10!white,
                title=Axioms for gr/1]
\begin{itemize}
    \item[4.] gr($c$) [if $c$ is a constant]
    \item[5.] $\textup{gr}(x_1) \land \ldots \land  \textup{gr}(x_m) \leftrightarrow \textup{gr}(f(x_1,\ldots,x_m))$ [$f$ is $m$-ary]
\end{itemize}
\end{tcolorbox}
Actually, LPTP deals with \emph{non-ground} terms and offers a predefined predicate  $gr/1$ that we can consider as a constraint.
This relation is useful for instance for 
dealing with negation as failure as
LPTP only allows negation by failure for \emph{ground} goals
(see the  definition $\textbf{T}  \texttt{\textbackslash +} G$).
%In this case, assuming termination, negation by failure is equivalent to logical negation.
Back to our example, here is the FOF version:
\begin{verbatim}
fof(id4,axiom,gr('0')).
fof(id5,axiom,! [Xx6] : (gr(Xx6) <=> gr(s(Xx6)))).
\end{verbatim}
The ADD program contains two user-defined predicates, \texttt{add/3} and \texttt{nat/1}.
LPTP considers each user-defined predicate through three points of view: failure, success and termination.
So LPTP creates the following predicates: \texttt{add\_fails/3}, \texttt{add\_succeeds/3}, \texttt{add\_terminates/3}, 
and similarly for 
\texttt{nat/1}. These three viewpoints are linked with the following axioms,
where $R^s$ (resp. $R^f$ and $R^t$) denotes 
\texttt{R\_succeeds/3} (resp. \texttt{R\_fails/3} and \texttt{R\_terminates/3}). 
%$R\_{succeeds}$ (resp. $R\_\textit{fails}$ and $R\_{terminates}$).

%
\begin{tcolorbox}[coltitle=black!75!black, colbacktitle=black!10!white,
                title=Uniqueness axioms and totality axioms]
\begin{itemize}
    \item[6.] $\neg (R^s(\vv{x}) \land R^f(\vv{x}))$ [if $R$ is a user-defined predicate]
    \item[7.] $R^t(\vv{x}) \rightarrow (R^s(\vv{x}) \lor R^f(\vv{x}))$ [if $R$ is a user-defined predicate]
\end{itemize}
\end{tcolorbox}
Axiom 6 says that for any tuple of (possibly non-ground)  terms, we cannot have at the same time success and failure of $R$.
Axiom 7 states that given termination, we have success or failure. Altogether, it means that for any 
tuple of terms $\vv{x}$, assuming termination,
either $R(\vv{x})$ succeeds or (exclusively) $R(\vv{x})$  fails.
So for our example, we get:
\begin{verbatim}
fof(ida6,axiom,! [Xx7,Xx8,Xx9] : 
    ~ ((add_succeeds(Xx7,Xx8,Xx9) & add_fails(Xx7,Xx8,Xx9)))).
fof(ida7,axiom,! [Xx7,Xx8,Xx9] : 
    (add_terminates(Xx7,Xx8,Xx9) => 
    (add_succeeds(Xx7,Xx8,Xx9) | add_fails(Xx7,Xx8,Xx9)))).
fof(idn6,axiom,! [Xx10] : 
    ~ ((nat_succeeds(Xx10) & nat_fails(Xx10)))).
fof(idn7,axiom,! [Xx10] : 
    (nat_terminates(Xx10) => 
    (nat_succeeds(Xx10) | nat_fails(Xx10)))).
\end{verbatim}
\begin{tcolorbox}[coltitle=black!75!black, colbacktitle=black!10!white,
                title=Fixed point axioms for user-defined predicates $R$]
\begin{itemize}
    \item[8.] $R^s(\vv{x})  \leftrightarrow$ \textbf{S}$D^P_R(\vv{x})$, 
    \ $R^f(\vv{x})  \leftrightarrow$ \textbf{F}$D^P_R(\vv{x})$,
    \ $R^t(\vv{x})  \leftrightarrow$ \textbf{T}$D^P_R(\vv{x})$
\end{itemize}
\end{tcolorbox}
We recall that $D^P_R(\vv{x})$ denotes the definition of the completion \cite{Cla78} 
of the user-defined  procedure $R(\vv{x})$ in the logic program $P$.
In the previous section, we saw how to apply the operator \textbf{S}, \textbf{F} and \textbf{T}
to formulas. So for instance, the first equivalence $R^s(\vv{x})  \leftrightarrow$ \textbf{S}$D^P_R(\vv{x})$
defines $R^s(\vv{x})$. Back to our running example, we get:
\begin{verbatim}
fof(idns8,axiom,! [Xx1] : (nat_succeeds(Xx1) <=> 
    (? [Xx2] : (Xx1 = s(Xx2) & nat_succeeds(Xx2)) | Xx1 = '0'))).
fof(idnf8,axiom,! [Xx1] : (nat_fails(Xx1) <=> 
    (! [Xx2] : (~ (Xx1 = s(Xx2)) | 
                nat_fails(Xx2)) & ~ (Xx1 = '0')))).
fof(idnt8,axiom,! [Xx1] : (nat_terminates(Xx1) <=> 
    (! [Xx2] : ((~ (Xx1 = s(Xx2)) | nat_terminates(Xx2)))))).
\end{verbatim}    
and similarly for \texttt{add/3}.

Finally, for any  property
of the form $\forall \vv{x} [R^s(\vv{x}) \rightarrow \phi(\vv{x})]$, where $R(\vv{x})$ 
is a user-defined procedure  and  $\phi(\vv{x})$ an $\hat{\cal{L}}$ -formula,
we have an induction schema.
%Given a logic program $P$ as a finite text, there is a finite number
%of user-defined procedures but an infinite number of formula. 
The interactive prover LPTP is able to \emph{dynamically} generate 
an induction axiom on demand while the user interacts with it.
In our approach, we \emph{statically} generate  the induction axiom \emph{once} 
from the conjecture to be proved, if the conjecture can be easily rewritten as required.  
This is a potential source
of imprecision, but again we stay sound.
Let us examine a simple case. It is exactly what happens using LPTP,
which slightly generalizes \cite{Staerk98a}.
By \emph{directly recursive  user-defined predicate} in the box below, 
we forbid mutual recursive definitions.
Of course, LPTP is able to handle
mutually recursive properties, see \cite{Staerk95a} for some examples.
\begin{tcolorbox}[coltitle=black!75!black, colbacktitle=black!10!white,
                title=A (simplified) induction schema for a user-defined predicate $R$ ]
    Let $R$ be a directly recursive  user-defined predicate and 
    let $\phi(\vv{x})$ be an $\hat{\mathcal{L}}$-formula such that the length of $\vv{x}$ is equal to the arity of $R$.\\
    Let $sub(\phi(\vv{x})/R)$ be the formula to be proven
    $\forall \vv{x} (R^s(\vv{x}) \rightarrow \phi(\vv{x}))$. \\
    Let $closed(\phi(\vv{x})/R)$ be the formula obtained from 
    $\forall \vv{x} (\textbf{S} D^P_R(\vv{x}) \rightarrow  R^s(\vv{x}))$ by replacing 
    \begin{itemize}
    \item $R^s(\vv{x})$ by $\phi(\vv{x})$  	on the right of $\rightarrow$,
    \item all occurrences of $R(\vv{t})$ appearing on the left of $\rightarrow$ by $\phi(\vv{t}) \land R(\vv{t})$.
    \end{itemize}
    Then the induction axiom is the following formula:
\begin{itemize}
    \item[9.] $closed(\phi(\vv{x})/R) \rightarrow sub(\phi(\vv{x})/R)$
\end{itemize}
\end{tcolorbox}
Let us apply this axiom to the following property, informally stated as: 
for any term $x$, if \texttt{nat($x$)} then \texttt{add($x$,0,$x$)}.
Expressed in LPTP, it gives: 
for any term $x$, 
if \texttt{nat\_succeeds($x$)} then \texttt{add\_}  \texttt{succeeds($x$,0,$x$)}, 
which is exactly 
%In FOF syntax:
%\texttt{! [Xx] : (nat\_succeeds(Xx) => add\_succeeds(Xx,'0',Xx)))}
the formula $sub(\phi(\vv{x})/R)$ of axiom 9. 
So $R \equiv \texttt{nat}$,  $R^s \equiv$ \texttt{nat\_} \texttt{succeeds}
and $\phi(\vv{x}) \equiv \texttt{add\_succeeds}(x,0,x)$.

For the left-hand side of axiom 9,   we start from
$$\forall x (\textbf{S} D^{ADD}_{nat}(x)\rightarrow \texttt{nat\_succeeds}(x))$$
%From the program ADD, 
We have $D^{ADD}_{nat}(x) \equiv x=0 \lor \exists y (x = s(y) \land \texttt{nat}(y))$.
We replace $\texttt{nat}(y)$ by $\texttt{nat}(y) \land \texttt{add\_succeeds}(y,0,y)$.
We replace $\texttt{nat\_succeeds}(x)$ by $\texttt{add\_succeeds}(x,$ $0,x)$. We get:
$\forall x \ (\textbf{S}$  
$[x=0 \lor \exists y (x = s(y) \land \texttt{nat}(y) \land $ $\texttt{add\_succeeds}(y,0,y))] 
\rightarrow \texttt{add\_succeeds}(x,0,x))$.
We apply \textbf{S} and obtain: $\forall x ( [x=0 \lor \exists y (x = s(y) \land \texttt{nat\_succeeds}(y) \land \texttt{add\_succeeds}(y,0,y))] 
\rightarrow \texttt{add\_succeeds}(x,0,x))$.

Summarizing, in FOF, associated with the property to be proved:
\begin{verbatim}
fof(lemma,conjecture,	
! [Xx] : (nat_succeeds(Xx) => add_succeeds(Xx,'0',Xx))).
\end{verbatim}
we obtain the following induction axiom:
\begin{verbatim}
fof(induction,axiom,(
    ! [Xx] : 
        ((? [Xx2] : (Xx = s(Xx2) & (nat_succeeds(Xx2) 
                                    & add_succeeds(Xx2,'0',Xx2))) 
        | Xx = '0') => add_succeeds(Xx,'0',Xx)) 
=> 
    ! [Xx] : (nat_succeeds(Xx) => add_succeeds(Xx,'0',Xx)))).
\end{verbatim}
%
%By copy/pasting all the FOF formulas of this section to the TPTP web site, one can check 
%that the syntax is correct and try to prove it with some selected provers. 
%
We can gather all the 15 axioms, including the axioms defining
\texttt{add\_success/3}, \texttt{add\_fails/3}, and \texttt{add\_terminates/3}
and the conjecture plus its induction axiom
in a file, say \texttt{test.fof} and submit it
%\footnote{E.g., in the terminal  \texttt{eprover --auto test.fof} and  \texttt{vampire test.fof}} 
to the E prover or to Vampire. Both systems
will find a refutation in a fraction of 
a second %\footnote{0.05s for Vampire, 0.017s for E on a MacBook Air, Apple M2, 16Go, macOS Sonoma.}
on a standard laptop.

It allows us to conclude  
for any term $x$, if \texttt{nat($x$)} then \texttt{add($x$,0,$x$)}
is true. Operationally, for any natural number $n$, 
in the Prolog search tree corresponding to the goal
$\texttt{add}(s^n(0),0,s^n(0))$, the empty clause appears.
Assuming termination, which will be shown later,
it means that the user will get  (at least) one positive answer for the query
$\texttt{:- add}(s^n(0),0,s^n(0)).$ when executed with any ISO-Prolog system.

Here's the manual proof of the same property in 
its LPTP version (a Prolog file), 
followed by 
its \TeX \ version produced by LPTP.
Using the interactive LPTP Emacs mode, we began this proof by invoking the \texttt{ind} tactic, asking for an inductive
proof. Both the base case and the inductive case were 
automatically generated and completed by LPTP.
\begin{verbatim}
:- lemma(add:x_0_x, all [x]: succeeds nat(?x) => succeeds add(?x,0,?x),
induction([all x: succeeds nat(?x) => succeeds add(?x,0,?x)],
 [step([],[],[],succeeds add(0,0,0)),
  step([x], [succeeds add(?x,0,?x), succeeds nat(?x)], [], 
   succeeds add(s(?x),0,s(?x)))])).
\end{verbatim}

%TeX proof.
\begin{plain}
%\input{add-x0x.tex}	
\input{add-x0x.tex}	
\end{plain}

\subsection{A second property}
Now let us consider the following property:
for any $x$, $y$ and $z$ such that $\texttt{nat}(x)$, $\texttt{nat}(y)$ and $\texttt{add}(s(x),y,z)$, we have 
 $\texttt{add}(x,s(y),z)$.
Let us first assert the previous property as an axiom,
which can now be freely used by the automated prover, then we have our new conjecture:
 \begin{verbatim}
fof('lemma-(add:x_0_x)',axiom,
  ! [Xx] : (nat_succeeds(Xx) => add_succeeds(Xx,'0',Xx))).

fof('lemma-(add:succ)',conjecture,
  ! [Xx,Xy,Xz] : (((nat_succeeds(Xx) & nat_succeeds(Xy)) 
                    & add_succeeds(s(Xx),Xy,Xz)) 
                    => add_succeeds(Xx,s(Xy),Xz))).
\end{verbatim}

In order to generate an induction axiom for this property,
we first rewrite it in the form $\forall \vv{x} [R^s(\vv{x}) \rightarrow \phi(\vv{x})]$
and we apply the simplified induction schema for user-defined predicates.
It gives:
\begin{verbatim}
fof(induction,axiom,(
! [Xx] : 
    ((? [Xy25] : 
      (Xx = s(Xy25) & (nat_succeeds(Xy25) 
       & ! [Xy,Xz] : ((add_succeeds(s(Xy25),Xy,Xz) 
                       & nat_succeeds(Xy)) 
      => add_succeeds(Xy25,s(Xy),Xz)))) 
    | Xx = '0') => 
      ! [Xy,Xz] : ((add_succeeds(s(Xx),Xy,Xz) & nat_succeeds(Xy)) 
         => add_succeeds(Xx,s(Xy),Xz))) 
 => ! [Xx] : (nat_succeeds(Xx) 
    => ! [Xy,Xz] : ((add_succeeds(s(Xx),Xy,Xz) & nat_succeeds(Xy)) 
                    => add_succeeds(Xx,s(Xy),Xz))))).
 \end{verbatim}

Again, we can gather all axioms, the conjecture and its induction axiom
in a file and submit it to Vampire,
which will find a refutation in about one minute.

%Here's a manual LPTP proof of the same property in its \TeX \ version.
%We began this proof by invoking the \texttt{ind} tactic, asking for an inductive
%proof. Both the base case and the inductive case were 
% manually generated and  completed. 
%Clearly, the proof is more complex than the previous one.
%
%\begin{plain}
%\input{add_succ.tex}	
%\end{plain}

 \subsection{Commutativity of Peano addition}
We are now equipped to consider commutativity of Peano addition: 
for any $x,y,z$, if $\texttt{add}(x,y,z)$ then  $\texttt{add}(y,x,z)$.
Of course, stated this way, the property is false.
% and
%Prolog can confirm this:
%\begin{verbatim}
%?- add(s(s(0)),c,s(s(c))).
%true.
%?- add(c,s(s(0)),s(s(c))).
%false.
%?- 
%\end{verbatim}
We need to enforce that $x$ and $y$ are Peano numbers.
%Hence $z$ is be a Peano number and commutativity holds.
%
So first we add our two previous properties as axioms.
Here is our new conjecture, associated with its induction axiom:
\begin{verbatim}
fof('theorem-(add:commutative)',conjecture,
  ! [Xx,Xy,Xz] : (((nat_succeeds(Xx) & nat_succeeds(Xy)) 
                   & add_succeeds(Xx,Xy,Xz)) 
                  => add_succeeds(Xy,Xx,Xz))).	

fof(induction,axiom,
(! [Xx] : 
  ((? [Xy26] : (Xx = s(Xy26) & (nat_succeeds(Xy26) 
    & ! [Xy,Xz] : ((add_succeeds(Xy26,Xy,Xz) & nat_succeeds(Xy)) 
    => add_succeeds(Xy,Xy26,Xz)))) 
    | Xx = '0') => 
      ! [Xy,Xz] : ((add_succeeds(Xx,Xy,Xz) & nat_succeeds(Xy)) 
    => add_succeeds(Xy,Xx,Xz))) 
=> 
! [Xx] : (nat_succeeds(Xx) => 
  ! [Xy,Xz] : ((add_succeeds(Xx,Xy,Xz) & nat_succeeds(Xy)) 
  => add_succeeds(Xy,Xx,Xz))))).
\end{verbatim}

The conjecture is proved in a fraction of  a second by Vampire. 
%The  LPTP proof  explicitly uses the two previous lemmas:
%
%\begin{plain}
%\input{add_comm.tex}	
%\end{plain}

\subsection{Some termination proofs}
%\label{termination}
Finally, let us prove some termination properties about \texttt{add/3}. 
It is immediate to see that the Prolog proof of
$\texttt{add}(x,y,z)$ terminates if $\texttt{nat}(x)$ or $\texttt{nat}(z)$. 
We  prove this by stating two lemmas which we  will gather  in a theorem. 
Here are the LPTP properties and their proofs (we omit the second one).

%
%Tex Proof.
\begin{plain}

\input{add-term.tex}	
\end{plain}
Each of the three statements is proved in a fraction %\footnote{Less than 0.1s.} 
of a second by Vampire.
Our compiler generates an instance of the induction axiom for each lemma and not for the theorem.
For instance, here is the first conjecture and its induction axiom:
\begin{verbatim}
fof('lemma-(add:term:1)',conjecture,
  ! [Xx,Xy,Xz] : (nat_succeeds(Xx) => add_terminates(Xx,Xy,Xz))).

fof(induction,axiom,(
! [Xx] : 
  ((? [Xx2] : (Xx = s(Xx2) & (nat_succeeds(Xx2) 
                    & ! [Xy,Xz] : add_terminates(Xx2,Xy,Xz))) 
    | Xx = '0') 
  => ! [Xy,Xz] : add_terminates(Xx,Xy,Xz)) 
=> 
! [Xx] : (nat_succeeds(Xx) => ! [Xy,Xz] : add_terminates(Xx,Xy,Xz)))).
\end{verbatim}

\section{Experimental Results}
\label{experimental:results}

We applied the schema explained in the previous sections to various libraries 
available with LPTP which we summarize now. 
The library \texttt{nat} defines some basic Peano relations
with the expected properties. 
The library \texttt{ack} defines the relational version of the Ackermann 
function with three properties (see below).
The library \texttt{gcd} defines a version of the greatest common divisor relation, 
with its full correctness proof. 
The library \texttt{int} defines integers.
The library \texttt{list}  proposes some elementary relations
about lists with their properties. The library \texttt{suffix} defines two versions of the sublist relation,
one as the prefix of a suffix, the other as the suffix of a prefix,  and shows that the two versions are equivalent
w.r.t. termination, success and failure. Similarly, the library \texttt{reverse} defines the two classical versions
of the reverse relation, one with the append relation, the other with an accumulator and shows their full equivalence.
The library \texttt{permutation} defines the permutation relation with 
some useful properties  for the correctness proofs of the sorting algorithms defined 
in the libraries \texttt{sort} and \texttt{mergesort}. The library \texttt{taut} 
defines a tautology checker for propositional formulas, 
together with its correctness proof (see \cite{Staerk96a} for a detailed description).

How do we process such files? Given a program from the LPTP  library,  
we first enumerate the requirements for trying to prove the properties listed in its associated LPTP proof file.  
Requirements are the logic program $P$ and the associated LPTP proof file. 
If $P$ depends on other logic programs, we must include them.
If the associated LPTP proof file uses other proof files, 
we must include them as well. 
We assume  there is no circularity such as assuming a lemma before trying to prove it.
We use these requirements to build a target logic program $P'$ and a target LPTP proof file.  
Then $P'$ is compiled into the FOF version of IND($P'$).  
Each fact (i.e., lemma, corollary or theorem) %to be proved automatically from the target proof file 
is compiled as a FOF conjecture (possibly with its induction axiom) and stored in a single file. 
Such file also contains the logic theory IND($P'$) compiled as FOF axioms.  
Previously processed FOF conjectures  are converted as FOF axioms as well.
As a result, we produce as many FOF files as there are facts in the initial LPTP proof file.
%conjectures to prove.  
At last, both the E Theorem Prover and Vampire are applied to each FOF file with the commands
\texttt{vampire --mode casc -m 16384 --cores 7 -t \$TO \$FILE} and
\texttt{eprover} \texttt{--delete-bad-limit=2000000000 --definitional-cnf -s --auto-schedule=8 --proof-object --cpu-limit=\$TO \$FILE}.

%This experiment can be reproduced for all the programs in the src directory or each of them separately.  
%Tests have been conducted on a MacBook Pro, 2022 (8 cores), Apple M2, 24 GB, macOS Sonoma.

We gather the results in Table \ref{tab:successrate}. The first column gives the library names.
The second column gives the number of (lemmas/corollaries/theorems) 
of the associated proof file.
The remaining nine columns can be divided in three  groups.
On a MacBook Pro, 8 cores, M2, 24 GB, macOS Sonoma,
the first group gives the success rate for a 1 second timeout
for the E prover (column E-1s), Vampire (column V-1s) and for the combination of the two provers (column EV-1s).
The second group (resp. third group) gives the success rate  for a timeout of 10 seconds (resp. 60 seconds).

%\begin{center}
%\begin{tabular}{lcrrrrrrrrr}
%\hline
%FOL & FOF & FOL & FOF  \\
%\hline
%$A \land B$ &  \texttt{A  \& B} & $\neg p(x)$ & \texttt{\~ \ p(X)} \\
%$A \lor B$ & \texttt{A | B} & $\exists x . A$ & \texttt{?[X] : A}\\
%$A \rightarrow B$ & \texttt{A => B} & $\forall x . A$ & \texttt{![X] : A}\\
%\hline
%\end{tabular}
%\end{center}

\begin{table}[h!]
\begin{center}
\begin{tabular}{| l | c | rrrrrrrrr |}
\hline
\textit{lib}&{\#}&E-1s&  V-1s&EV-1s&  E-10s&  V-10s&  EV-10s&  E-60s&  V-60s& EV-60s\\
\hline
\hline
nat &91 &70\% &88\% &88\% &76\% &95\% &95\% &78\% &97\% &97\%\\ 
\hline
gcd &11 &45\% &45\% &45\% &45\% &45\% &45\% &45\% &45\% &45\%\\ 
\hline
ack &3 &33\% &33\% &33\% &33\% &33\% &33\% &33\% &33\% &33\%\\
\hline
int &67 &76\% &82\% &87\% &79\% &88\% &90\% &79\% &91\% &91\%\\ 
\hline
list &84 &75\% &94\% &94\% &80\% &96\% &96\% &81\% &99\% &99\%\\ 
\hline
suffix &31 &94\% &100\% &100\% &94\% &100\% &100\% &97\% &100\% &100\%\\ 
\hline
reverse &25 &72\% &88\% &88\% &84\% &100\% &100\% &84\% &100\% &100\%\\ 
\hline
permut. &42 &48\% &71\% &71\% &60\% &79\% &81\% &62\% &86\% &86\%\\ 
\hline
sort &42 &45\% &62\% &62\% &50\% &74\% &74\% &55\% &76\% &76\%\\ 
\hline
merges. &24 &79\% &88\% &88\% &79\% &92\% &92\% &79\% &100\% &100\%\\ 
\hline
taut &43 &65\% &81\% &81\% &70\% &84\% &84\% &74\% &84\% &84\%\\
\hline
\end{tabular}
\caption{Experimental Evaluation}
\label{tab:successrate}
\end{center}
\end{table}

%
% ?- SR is (97+45+33+91+99+100+100+86+76+100+84)/11.
% SR = 82.81818181818181.
%
Let us comment these results.
The \texttt{gcd} Prolog file contains a mutually recursive  definition 
for the predicates \texttt{gcd/3} and \texttt{gcd\_leq/3}. Proving properties
of such definitions is  currently out of scope of our translation schema.

The \texttt{ack} proof file contains the following three properties.
The first one is successfully checked. The last two ones
cannot be proved with our simplified induction schema. Indeed,
the LPTP proofs use an induction inside the top level induction,
which is  out of scope of our translation schema.

%Tex Proof.
\begin{plain}
\input{ack.tex}	
\end{plain}

\section{Related Work}

There is quite a few Prolog verification frameworks, see e.g.  \cite{Deransart93a,FerrandD93,Apt94b,PedreschiR99}
and more recently  \cite{Drabent16a}.
Most of them aim at \emph{paper and pencil} proofs. 
Although they may offer interesting and elegant methods, 
the validity of the proofs relies on the usual mathematical writing in natural language, 
and proofs are not automatically  checked.  
In our opinion, writing and verifying such hand-written proofs
can be a time consuming  and error-prone 
process compared to  a push-button approach as the one we present here.

Recently, some quite interesting works have been reported
on including datatypes, taking into account the acyclicity of their values, 
and induction in modern first-oder theorem provers, see, e.g.,
\cite{BlanchettePR18, HajduHKV21}.  We have not yet tested these extensions
within our framework.

For Answer Set Programming (a declarative specification language with a Prolog syntax, 
oriented towards knowledge representation and search problems),
\cite{FandinnoLLS20}  describes  an approach toward verification 
in which Vampire  checks the equivalence of Answer Set programs.

Some programming languages include automated verification tools \emph{by design}.
For example, Dafny \cite{Leino12} makes heavy use of SMT solving. 
The Why3  system \cite{FilliatreP13} allows to export verification conditions
to many automatic and interactive theorem provers.

An earlier account of the integration of automated 
and interactive theorem proving is described in \cite{Ahrendt98}.
As already announced in the introduction, 
most interactive theorem provers now include 
the possibility to run some automated theorem provers.
Starting with Isabelle, \cite{MengP04,PaulsonB10,BlanchetteKPU16,Paulson22a},
\emph{hammers} %\footnote{See also \cite{Paulson22a} for a historical account about Sledgehammer.}  
can  be found in e.g., 
ACL2, \cite{JoostenKU14},  Coq, \cite{CzajkaEK18} and Lean, \cite{LeanHammer19}.

\section{Conclusion}

Let us recall the questions of the introduction
and propose our answers after this %first %initial 
experiment:
%Here are our answers to the questions of the introduction
%after this initial experiment:
\begin{itemize}
	\item Can we also use the TPTP FOF \emph{Esperanto} to formulate
the logic theory St{\"a}rk associates to a logic program? Yes. 
One axiom schema was not implemented: Axiom 3 which forbids rational terms.
Another one was partially implemented:  Axiom 9 for induction. Actually an inductive
argument inside an inductive proof is not possible with our approach.
We lose precision but in both cases we stay sound.

\item Can we use \emph{off-the-shelf} TPTP provers and obtain automatic proofs in 
reasonable time? Yes.
We use Vampire and the E prover with their most basic options, essentially a timeout.
Although Vampire seems to find refutations faster,
the E prover can sometimes find proofs while Vampire cannot conclude within the time limit.
Hence the two provers are complementary. For the moment, we did not try %the
advanced features offered by the provers 
like the one proposed in \cite{KovacsRV17}.
% for directly dealing with finite trees. 

\item Can we get  an acceptable success rate with such an approach? Yes.
With the E prover and Vampire running in parallel, the average success rate we get
from our benchmark is about 83\% for a one minute timeout on a standard laptop.
\end{itemize}

Compared to the efforts one spends while manually, laboriously elaborating certain proofs
with an interactive theorem prover,
the use of state of the art automated theorem provers is clearly a time-saver. 
We did not expect such a good success rate for this first experiment.
We think there are various reasons that can explain it. 
Clearly, the computing power of our current laptops is huge 
and automated theorem provers have been largely improved.
Also, thanks to St{\"a}rk's ideas, 
the clean and simple semantics of both the pure subset of Prolog targeted by LPTP and the LPTP specification language --  essentially first-order logic -- 
implies a straightforward translation to FOF.
Last but not least, St{\"a}rk's art of proving, 
by slicing the proofs of  the LPTP library properties 
into manageable lemmas,  certainly has an  impact on the success rate we obtain.

Finally, there is room for improvement of the presented work, which
can be considered as a first approach towards a hammer for LPTP
according to \cite{BlanchetteKPU16}. 
The first step  of a hammer -- the \emph{premise selector},
which  selects subparts of the LPTP library potentially useful for a proof --
and the third step -- the \emph{proof reconstruction module}, which  rewrites the proof
found by the automatic prover in the LPTP proof format 
-- are yet to be investigated.

\vspace{0.5cm}
\noindent
\textbf{Acknowledgements.} We thank 
Manuel Hermenegildo, Daniel Jurjo, Pedro L\'opez-Garcia, and Jose Morales
for stimulating discussions about  LPTP, 
Geoff Sutcliffe for his help with TPTP and the reviewers for their constructive comments.

%\newpage
% \bibliographystyle{eptcs}
% \bibliography{biblio.bib}

\input{bbl}
%\section*{Appendix}
%The present paper already appeared 
%in the 2024 LPAR \emph{Complementary} Volume,
%which is not considered as the official LPAR proceedings by its editors.
%According to the editors, it does not prevent submission to an other conference.

\end{document}

%% file: add-x0x.tex
\input proofmacros.tex

!lemma{add:x\string_0\string_x}{!0!1!all[x]!,!1(!1!S !Tt{nat}(x)!2!to !1!S !Tt{add}(x,!Tt{0},x)!2)!2!2}

!Pr{
!Ind{$!0!1!all[x]!,!1(!1!S !Tt{nat}(x)!2!to !1!S !Tt{add}(x,!Tt{0},x)!2)!2!2$}
!Stp{none}
!Estp{!0!1!S !Tt{add}(!Tt{0},!Tt{0},!Tt{0})!2}
!Stp{$!0!1!S !Tt{add}(x,!Tt{0},x)!2$ and $!0!1!S !Tt{nat}(x)!2$}
!Estp{!0!1!S !Tt{add}(!Tt{s}(x),!Tt{0},!Tt{s}(x))!2}
!Eind}
!Epr

!end

%% file: proofmacros.tex
%   Author: Robert Staerk <staerk@math.stanford.edu>
%  Created: January 1995
%  Updated: Fri Mar 12 09:05:39 2004 
% Filename: proofmacros.tex
%
% Font for operators `succeeds', `fails' and `terminates'.
%
\font\bfsf=cmssbx10
%
% The style of proofs is "ragged right".
%
\rightskip 0pt plus 10cm
\tolerance=400
%\hoffset=-4.5mm
%
% Macros used in formulas:
%
%  \D          definition
%  \F          failure operator
%  \It          italic font for defined functions
%  \S          success operator
%  \T          termination operator
%  \Tt         typewriter font for predicates
%  \all        universal quantifier
%  \app        X ** Y
%  \eq         equality
%  \eqv        equivalence
%  \ex         existential quantifier
%  \is         X is Y
%  \land       conjunction
%  \leq        less than or equal to
%  \lnot       negation
%  \lor        disjunction
%  \neq        disequality
%  \sub        subset
%  \to         implication
%  \v          integer variables
%
\def\D{\mathop{\hbox{\bfsf D}}}
\def\F{\mathop{\hbox{\bfsf F}}}
\def\It#1{\hbox{\it #1}}
\def\S{\mathop{\hbox{\bfsf S}}}
\def\Tt#1{\hbox{\tt #1}}
\def\T{\mathop{\hbox{\bfsf T}}}
\def\all[#1]{\forall #1}
\def\app{\nobreak\mathbin{**}\nobreak}
\def\eqv{\leftrightarrow\penalty\levcount}
\def\eq{\nobreak=\nobreak}
\def\ex[#1]{\exists #1}
\def\is{\nobreak\mathbin{\hbox{\tt is}}\nobreak}
\def\land{\wedge\penalty\levcount}
\def\lor{\vee\penalty\levcount}
\def\sub{\nobreak\subseteq\nobreak}
\def\to{\rightarrow\penalty\levcount}
\def\apply{\nobreak\mathbin{/. }\nobreak}
\def\v#1{v_{#1}}
%
% The depth of formulas:
%
\newcount\levcount
\def\0{\global\levcount=20}
\def\1{\global\advance\levcount by 20}
\def\2{\global\advance\levcount by -20}
%
% Underscores (cf. ^ and ^^ in manmac.tex):
%
\newif\ifref
\reffalse
%\def\specialunderscore{\ifmmode_\else{\tt\char"5F}\fi} 
%\def\specialunderscore{\ifmmode_\else\ifref-\else\string_\fi\fi}
%\def\specialunderscore{\ifmmode_\else\ifref-\else\textunderscore\fi\fi}
%\def\specialunderscore{\_}
% FM: original code, does not work
%\def\specialunderscore{\ifmmode_\else\ifref-\else\_\fi\fi}
%\catcode`\_=13 % active
%\let _=\specialunderscore
%
% Labels and backward references:
%
\def\label#1#2{\reftrue\expandafter\edef\csname#1:#2\endcsname{\Hlink{\number\thmcount}}%
\Htarget{\number\thmcount}%
\edef\next{\write\auxout{\string\newlabel{#1}{#2}{\jobname}{\number\thmcount}}}%
\next\reffalse\ignorespaces}
\def\by#1#2{\penalty 20\ by\penalty 20\ #1~\reftrue%
\expandafter\ifx\csname#1:#2\endcsname\relax#2%
\else\csname#1:#2\endcsname\fi\reffalse~[{\it #2}]}
\def\newlabel#1#2#3#4{\expandafter\edef\csname#1:#2\endcsname{#4 in Module {\tt #3}}}
%
% Hyperlinks (based on `hyperref.dtx' by Sebastian Rahtz):
%
\def\Hlink#1{#1}
\def\Htarget#1{}
%
% Uncomment the following lines if you want to use hyperlinks.
%
\def\Hend{}
\edef\Hhash{\string#}
\edef\Hquote{\string"}
\def\Hhref#1{}
\def\Hname#1{}
\def\Hlink#1{\Hhref{#1}#1\Hend}
\def\Htarget#1{\Hname{#1}\Hend}
%
% Theorem, Lemma, Corollary, Definition, Axiom:
%
\newcount\thmcount
\thmcount=0
\newcount\indcount
\def\inc{\global\advance\indcount by 1\hangindent\indcount em}
\def\dec{\global\advance\indcount by-1}
\def\nl{\par\hangindent\indcount em\noindent\kern\indcount em\ignorespaces} 
\def\lev{$\strut_{\number\indcount}$}
\def\Module#1#2#3{\bigskip\goodbreak % \vfil\allowbreak\vfilneg
  \advance\thmcount by1\label{#1}{#2}
%  \noindent{\bf #1~\the\thmcount}~[{\it #2}] $#3$.\allowbreak}
  \noindent{\bf #1}~[{\it #2}] $#3$.\allowbreak}
\def\theorem{\Module{Theorem}}
\def\lemma{\Module{Lemma}}
\def\corollary{\Module{Corollary}}
\def\definition{\Module{Definition}}
\def\axiom{\Module{Axiom}}
%
% Proof steps:
%
\def\Pr{\allowbreak\smallskip\noindent\global\indcount=0{\bf Proof. }\nobreak}
\def\Epr{\hbox{\rlap{$\sqcup$}$\sqcap$}\smallskip\allowbreak}
\def\Ass#1{\nl Assumption\lev: $#1$. \inc}
\def\Eass#1{\dec\nl Thus\lev: $#1$.}
\def\Cas#1{\nl Case\lev: $#1$. \inc}
\def\Ecas{\dec}
\def\Fin#1{\nl Hence\lev, in all cases: $#1$.}
\def\Dir#1{\nl Indirect\lev: $#1$. \inc}
\def\Edir#1{\dec\nl Thus\lev: $#1$.}
\def\Con#1{\nl Contra\lev: $#1$. \inc}
\def\Econ#1{\dec\nl Thus\lev: $#1$.}
\def\Ex[#1]#2{\nl Let\lev\ $#1$ with $#2$. \inc}
\def\Eex#1{\dec\nl Thus\lev: $#1$.}
\def\Ind#1{\nl Induction\lev: #1. \inc}
\def\Eind{\dec}
\def\Stp#1{\nl Hypothesis\lev: #1. \inc}
\def\Estp#1{\dec\nl Conclusion\lev: $#1$.}
\def\noproofs{\let\Pr=\nil\let\Epr=\par}
\def\nil#1{}
% FM
%\def\title#1{\noindent{\bf File:} {\tt#1.pr}\par}
%
% .aux
%
\newread\inputcheck
\def\inputaux#1{\openin\inputcheck #1 \ifeof\inputcheck \message
  {No file #1.}\else\closein\inputcheck \relax\input #1 \fi}
\newwrite\auxout
\openout\auxout=\jobname.aux
\let\endsave=\end
\def\end{\write\auxout{}\closeout\auxout\endsave}
%
%
% "!" is escape character like backslash "\"
%
\catcode `!=0
\catcode `@=11
\catcode `#=11
\catcode `&=11
%
%\noproofs
%
%
% End of proofmacros.tex

%% file: add-term.tex
\input proofmacros.tex
!lemma{add:term:1}{!0!1!all[x,y,z]!,!1(!1!S !Tt{nat}(x)!2!to !1!T !Tt{add}(x,y,z)!2)!2!2}
!Pr{
!Ind{$!0!1!all[x]!,!1(!1!S !Tt{nat}(x)!2!to !1!all[y,z]!,!1!T !Tt{add}(x,y,z)!2!2)!2!2$}
!Stp{none}
!Estp{!0!1!all[y,z]!,!1!T !Tt{add}(!Tt{0},y,z)!2!2}
!Stp{$!0!1!all[y,z]!,!1!T !Tt{add}(x,y,z)!2!2$ and $!0!1!S !Tt{nat}(x)!2$}
!Estp{!0!1!all[y,z]!,!1!T !Tt{add}(!Tt{s}(x),y,z)!2!2}
!Eind}
!Epr

!lemma{add:term:3}{!0!1!all[x,y,z]!,!1(!1!S !Tt{nat}(z)!2!to !1!T !Tt{add}(x,y,z)!2)!2!2}
!Pr{Similar.}
!Epr

!theorem{add:term}{!0!1!all[x,y,z]!,!1(!1!1!S !Tt{nat}(x)!2!lor !1!S !Tt{nat}(z)!2!2!to !1!T !Tt{add}(x,y,z)!2)!2!2}
!Pr{
!Ass{!0!1!1!S !Tt{nat}(x)!2!lor !1!S !Tt{nat}(z)!2!2}
!Cas{!0!1!S !Tt{nat}(x)!2}
$!0!1!T !Tt{add}(x,y,z)!2$!by{Lemma}{add:term:1}.
!Ecas
!Cas{!0!1!S !Tt{nat}(z)!2}
$!0!1!T !Tt{add}(x,y,z)!2$!by{Lemma}{add:term:3}.
!Ecas
!Fin{!0!1!T !Tt{add}(x,y,z)!2}
!Eass{!0!1!1!1!S !Tt{nat}(x)!2!lor !1!S !Tt{nat}(z)!2!2!to !1!T !Tt{add}(x,y,z)!2!2}}
!Epr

!end

!end

%% file: ack.tex
\input proofmacros.tex
!lemma{ackermann:types}{!0!1!all[m,n,k]!,!1(!1!1!S !Tt{ackermann}(m,n,k)!2!land !1!S !Tt{nat}(n)!2!2!to !1!S !Tt{nat}(k)!2)!2!2}
!lemma{ack:existence}{!0!1!all[m,n]!,!1(!1!1!S !Tt{nat}(m)!2!land !1!S !Tt{nat}(n)!2!2!to !1!ex[k]!,!1!S !Tt{ackermann}(m,n,k)!2!2)!2!2}
!lemma{ack:termination}{!0!1!all[m,n,k]!,!1(!1!1!S !Tt{nat}(m)!2!land !1!S !Tt{nat}(n)!2!2!to !1!T !Tt{ackermann}(m,n,k)!2)!2!2}
!end